\documentclass[english,pra, twocolumn]{revtex4-1}
\usepackage[T1]{fontenc}
\usepackage[latin9]{inputenc}
\usepackage{textcomp}
\usepackage{amsmath}
\usepackage{amssymb}
\usepackage{graphicx}

\makeatletter
\usepackage[colorlinks = true,
            linkcolor =red,
            urlcolor  =magenta,
            citecolor = blue,
            anchorcolor = blue]{hyperref}

\makeatother

\usepackage{babel}
\begin{document}
\title{General approach of weak-to-strong measurement transition for
    Fock-state-based pointer states}
\author{Yusuf Turek$^{1}$}
\email{yusuftu1984@hotmail.com}

\author{Janarbek Yuanbek$^{2}$}
\author{Ahmad Abliz$^{2}$}
\affiliation{$^{1}$School of Physics Science and Technology, Xinjiang University,
Urumqi, Xinjiang 830046, China}
\affiliation{$^{2}$School of Physics and Electronic Engineering, Xinjiang Normal
University, Urumqi, Xinjiang 830054, China}
\date{\today}
\date{\today}
\begin{abstract}
The transition from von Neumann\textquoteright s
projective strong measurement to Aharonov\textquoteright s weak measurement has recently received large attention, theoretical and experimental. In this work,
we present a general approach to describe the weak-to-strong measurement transition for Fock-state-based pointer states, and analyze in some details the case of coherent pointer states. A possible realization of our measurement scheme using trapped ions is also discussed.
\end{abstract}
\maketitle

\section{\label{sec:1}Introduction}
Measurements are at the basis of physics, and represent
the major tool for the quantitative understanding of our
world. In quantum systems, the measurement problem has fostered a better understanding of quantum theory itself, and its
applications to technology \citep{RevModPhys.75.715,Haroche2006,200Na,2013Na,2014Quantum,MATSUNO2017131,2020na,2021NA}.
In standard textbooks, quantum measurements are usually introduced using the projective measurement model formulated by
von Neumann \citep{Edwards1955}. The description of the Stern-Gerlach experiment is a typical example of
this kind of measurement \citep{Sakurai1995Modern}. This is a
{\em strong measurement} model where the strength of the interaction between the pointer (measuring device) system and the  measured one is strong enough to obtain the desired
information about the measured system after a single trial,
but the system itself collapses onto an eigenstate
of the measured system observable.
On the other hand, there is another type of measurement
modelm proposed in 1988 by Aharonov et al \citep{Aharonov1988}. In this scheme, which also follows the basic requirements of von Neumann measurement theory, the coupling between the measured and the measuring systems is weak. As a consequence, the initially prepared system state is not destroyed after a single trial,
and we still can get the required information statistically, after many repetitions \citep{Tollaksen_2010}.
In the {\em weak measurement} model, the value of the measured system observable is extracted in the form of {\em weak value} caused by the pre- and post-selection processes. This weak value is, in general, a complex number and different from the
eigenvalue or expectation value of the corresponding system
observable. Another important feature of weak values is that
they could exceed the range of the eigenvalues. This
feature is usually referred to as {\em weak signal amplification}
and has been found useful in the solution of several
problems in physics and related sciences \citep{Hosten787,PhysRevA.80.041803,PhysRevLett.102.173601,Lundeen2011,Pfeifer:11,Lundeen2012,aBoyd2013,boyd2013,Malik13,PhysRevA.88.053815,PhysRevLett.110.010404,PhysRevLett.123.150402,PhysRevA.100.032119,PhysRevLett.118.010402}.
For more details, the reader is referred to \citep{KOFMAN201243,Malik2014} and the references therein.

In a recent study \citep{PhysRevA.96.032114}, it has been shown that the nature of weak values is different from that of the expectation values of the system observable, though
weak values describe the interaction in the same way as the eigenvalue does. In general, a system observable is characterized by three sets of values: eigenvalues, (conditional) expectation values and weak values. The system information
we want to obtain from the pointer after measurement
is related to the pointer's shifts. In order to get these values,
we rely on different types of measurement models. The
eigenvalues and the expectation values of the system observable
are usually related to the position shift of the pointer after strong measurement. On the other hand, in the weak measurement procedures \citep{PhysRevA.76.044103}, the real and imaginary
parts of weak values are obtained from the position and momentum shifts of the pointer, respectively.
\par
As a matter of fact, a relevant point in von Neumann measurement model is that the coupling strength between the measured system
and the pointer may assume arbitrary values, so that we can
connect the weak and strong measurement schemes by adjusting it. Recently, the weak-to-strong measurement transition
problems has been investigated experimentally \citep{Nature2020} and theoretically
\citep{PhysRevA.103.052215,Yusuf2022} for some specific pointer states.
However, a general approach to the weak-to-strong measurement transition for arbitrary pointer states has not been yet investigated.

In this paper, we present a general approach to describe the weak-to-strong measurement transition problem for infinite-dimensional (pure) pointer states in the Fock space, i.e. for the so-called Fock-state-based pointer states. We know that
Fock number states $\vert n\rangle$ are the eigenstates
of the number operator $\hat{n}=\hat{a}^{\dagger}\hat{a}$ , i.e.,
$\hat{a}^{\dagger}\hat{a}\vert n\rangle=n\vert n\rangle$. Thus, we
can expand any quantum state $\vert\phi\rangle$ in the basis of $\vert n\rangle$
as $\vert\phi\rangle=\sum_{n}c_{n}\vert n\rangle$ with $c_{n}=\langle n\vert\phi\rangle$.
We consider the internal (e.g. spin or polarization)
and external degrees (e.g. spatial or field distribution) of freedom of the state as the measured system
and pointer (measuring device), respectively, and give the general
expressions of pointer shifts corresponding to the pointer
position and momentum operators after measurement. Since these expressions are derived without any approximation, they are
valid for any coupling, and we can described the transitions
from Aharonov\textquoteright s weak measurement to von Neumann\textquoteright s
projective strong measurement by tuning the coupling strength.

By using our general approach, the two extreme
cases can be recovered, in which the shifted values
of the pointers are directly associated with the
conditional expectation values (strong measurement regime) and
weak values (weak measurement regime), respectively. In particular, we derive a simple formula for the position shift of the pointer
in the weak measurement regime and verify its agreement with previous results. We also analyze in some details the case
of  coherent state pointers  by means of the Husimi-Kano $Q$
function in the phase space. Finally, we discuss the possible implementation of our transition measurement  model
in trapped ion systems.

The rest of this paper is organized as follows.
In Sec. \ref{sec:2}, we provide a brief introduction
to the von Neumann measurement by considering the weak
and strong interaction cases. In Sec. \ref{sec:3}, we derive
the final pointer state of our weak-to-strong measurement transition model and present the general
expressions of the position and momentum shifts. We also
illustrate how to realize the weak-to-strong measurement transition
by tuning the coupling strengths between the measured system and
the pointer. We also verify our results by comparison with
previous studies. In Sec.\ref{sec:3}, as an example of our general
approach, we investigate the coherent state based weak-to-strong measurement
transition. In Sec. \ref{sec:4}, we discuss the possible experimental
implementation of our model in trapped ion systems. In Sec. \ref{sec:5},
we provide a brief summary of this paper. Throughout this paper, we
use the unit $\hbar=1$.

\section{\label{sec:2} Brief introduction of quantum measurement }

The overall Hamiltonian of our system is the sum of three terms
$H=H_{s}+H_{p}+H_{int}$, where $H_{s}$ and $H_{p}$ denote the free
Hamiltonians of the measured system and the pointer (measuring
device), respectively, and $H_{int}$ is the interaction
Hamiltonian between the system and pointer. In the following,
since the readout information is related to the coupling between the system and the pointer, we describe the dynamics in an interaction picture and only consider the interaction
Hamiltonian $H_{int}$.
According to the von Neumann measurement theory, the general form
of the interaction Hamiltonian can be written as \citep{Edwards1955}
\begin{equation}
H_{int}=g(t)\hat{A}\otimes\hat{P},\text{\ \ \ensuremath{\int_{t_{0}}^{t}g(t)dt=g\delta(t-t_{0}).}}\label{eq:H-1}
\end{equation}
Here, $\hat{A}$ is the system observable that we want to
measure, whereas $\hat{P}$ represents the momentum operator
of the pointer. The position operator $\hat{X}$ is its canonically cojugated operator, i.e. $[\hat{X},\hat{P}]=i$.
The function $g(t)$ is  nonzero only in a finite interaction
time interval, and $g$ fully characterizes the coupling strength between the system and the pointer. If we denote by
$\vert a_{i}\rangle$ and eigenstate
of the observable $\hat{A}$ with the eigenvalue $a_{i}$, i.e., $\hat{A}\vert a_{j}\rangle=a_{j}\vert a_{j}\rangle$,
then the observable $\hat{A}$ can be written as \citep{Sakurai1995Modern}
\begin{equation}
\hat{A}=\sum_{j}a_{j}\vert a_{j}\rangle\langle a_{j}\vert.\label{eq:2}
\end{equation}
In general, given an observable $\hat{A}$, there are three sets
of relevant values to be considered: the eigenvalues, the (conditional) expectation values, and the weak values. According
to the measurement theory, these values can be obtained by a pointer's shift after measurement. Since each eigenvalue is a special case of a (conditional) expectation values and weak values of the observable, we now provide a brief review of the readout procedures to obtain (conditional) expectation values and weak values.

1.\textsl{ Expectation value}. Assume that the initial state
of the pointer is $\vert\Phi\rangle$ with wave function $\Phi(x)$,
and the measured system state is prepared in a superposition of eigenstates of $\hat{A}$, i.e., $\vert\psi_{i}\rangle=\sum_{j}\alpha_{j}\vert a_{j}\rangle$
with $\sum_{j}\vert\alpha_{j}\vert^{2}=\sum_{j}\vert\langle a_{j}\vert\psi_{j}\rangle\vert^{2}=1$.
Under the action of the unitary evolution operator $e^{-ig\hat{A}\otimes\hat{P}}$,
the total system state evolves into
\begin{align}
\vert\Psi\rangle & =e^{-ig\hat{A}\otimes\hat{P}}\vert\psi_{i}\rangle\otimes\vert\Phi(x)\rangle\nonumber \\
 & =\sum_{j}\alpha_{j}\vert a_{j}\rangle\otimes\vert\Phi(x-ga_{j})\rangle.\label{eq:3}
\end{align}
 The Eq. (\ref{eq:3}) implies that if a strong measurement of the observable $\hat{A}$ is performed with outcome $a_j$, then
the wave function of the pointer collapses onto $\Phi(x-ga_{j})$, i.e.  it is displaced by the amount $ga_{j}$ compared to the
initial wave function $\Phi(x)$. After some algebra, the final position
shift of the pointer for the strong measurement can be obtained as
\begin{align}
\delta x_{st} & ={\langle\Psi\vert\hat{X}\vert\Psi\rangle}-\langle\Phi\vert\hat{X}\vert\Phi\rangle\nonumber \\
 & =g\langle A\rangle.\label{eq:5-1}
\end{align}
Here, $\langle A\rangle$ is the expectation value of the observable
on the state $\vert\psi_{i}\rangle$, i.e.,
\begin{equation}
\langle A\rangle=\langle\psi_{i}\vert\hat{A}\vert\psi_{i}\rangle=\sum_{j}a_{j}\vert\alpha_{j}\vert^{2}.\label{eq:6-1}
\end{equation}
This result can be esily understood by writing the initial
density matrix
\begin{align}
\rho & =\vert\psi_{i}\rangle\langle\psi_{i}\vert\nonumber \\
 & =\sum_{j}\vert\alpha_{j}\vert^{2}\vert a_{j}\rangle\langle a_{j}\vert+\sum_{j,k}\alpha_{j}\alpha_{k}^{\ast}\vert a_{j}\rangle\langle a_{k}\vert.\label{eq:7}
\end{align}
It is clear that $\vert\alpha_{i}\vert^{2}$ is the probability
of eigenvalue $a_{i}$ of the observable $A$, corresponding to the
eigenstate $\vert a_{i}\rangle$, if the system is prepared in $\vert\psi_{i}\rangle$.
In order to obtain a determined outcome (i.e., an eigenvalue of the
measured observable corresponding to a given eigenstate ), the second
term (off-diagonal part which represented the coherence) of $\rho$
has to vanish after the measurement. This means that, after the measurement
the system is a mixture of the eigenstates of the measured observable,
i.e., $\sum_{j}P_{j}\rho P_{j}$, where $P_{j}=\vert a_{j}\rangle\langle a_{j}\vert$
are the projectors on the different eigenstates of $\hat{A}$. The disappearance of the off-dagonal terms in
the above mixture is usually referred to as the projection postulate.

2. \textsl{Conditional Expectation value}. In time-symmetric
quantum mechanics \citep{PhysRev.134.B1410,Aharonov_1991,A2008},
if we assume that a postselection onto the state $\vert\psi_{f}\rangle=\sum_{j}\beta_{j}\vert a_{j}\rangle$
of  $\vert\Psi\rangle$ {[}see Eq. (\ref{eq:3}){]}, is possible, than the unnormalized
final state of the total system is given by
\begin{equation}
\vert\Psi^{\prime}\rangle=\sum_{j}\alpha_{j}\beta_{j}^{\ast}\vert a_{j}\rangle\otimes\vert\Phi(x-ga_{i})\rangle\,,\label{eq:5}
\end{equation}
where $\sum_{j}\vert\beta_{j}\vert^{2}=\sum_{i}\vert\langle a_{i}\vert\text{\ensuremath{\psi_{f}}}\rangle\vert^{2}=1$.
In this case, after the measurement the position shift of the pointer
is proportional to the average value of the position operator
\begin{align}
\delta x_{st} & =\frac{\langle\Psi^{\prime}\vert\hat{X}\vert\Psi^{\prime}\rangle}{\langle\Psi^{\prime}\vert\Psi^{\prime}\rangle}-\langle\Phi\vert\hat{X}\vert\Phi\rangle=g\langle A\rangle_{c},
\end{align}
where $\langle A\rangle_{c}$ is the conditional expectation value
of the observable $A$, and defined as \citep{PhysRev.134.B1410}
\begin{align}
\langle A\rangle_{c} & =\frac{\sum_{j}a_{j}\vert\langle\psi_{f}\vert a_{j}\rangle\langle a_{j}\vert\psi_{i}\rangle\vert^{2}}{\sum_{k}\vert\langle\psi_{f}\vert a_{k}\rangle\langle a_{k}\vert\psi_{i}\rangle\vert^{2}}\nonumber \\
 & =\frac{\sum_{j}a_{j}\vert\alpha_{j}\beta_{j}^{\ast}\vert^{2}}{\sum_{j}\vert\alpha_{j}\beta_{j}\vert^{2}}.\label{eq:6}
\end{align}
This value is also referred to a the postselected strong
value of $\hat{A}$.
In the above derivations, we have assumed that the coupling
between the measured system and the pointer is strong enough
to make the sub-wave-packets of the pointer corresponding to  different eigenvalues distinguishable, i.e., $g\triangle a\gg\sigma$, where $\triangle a=a_{j}-a_{j-1}$
and $\sigma$ represent the differences of neighboring eigenvalues
and the width of the sub-wave-packets, respectively.

3. \textsl{Weak value}. At variance with the situation considered above,
if $g\triangle a\text{\ensuremath{<\sigma}}$, i.e., if system-pointer
coupling is sufficiently weak, we cannot get the
needed information after a single trial. In this case, we may
consider the expansion of the unitary operator up to the first order
term, such that the first line of Eq. (\ref{eq:3}) may be written as
\begin{equation}
\vert\eta\rangle\approx(1-ig\hat{A}\otimes\hat{P})\vert\psi_{i}\rangle\otimes\vert\Phi(x)\rangle.
\end{equation}
As for the conditional strong measurement process, if we take a
postselection on the state $\vert\psi_{f}\rangle$, then the
(unnormalized) total system state  may be written
\begin{align}
\vert\eta^{\prime}\rangle & \approx e^{-ig\langle A\rangle_{w}\hat{P}}\vert\psi_{f}\rangle\otimes\vert\Phi(x)\rangle\nonumber \\
 & \propto\vert\psi_{f}\rangle\otimes\left|\Phi\left(x-gtRe\left[\langle A\rangle_{w}\right]\right)\right\rangle .\label{eq:11}
\end{align}
 After the measurement, the position and momentum shifts of the pointer
are given by \citep{PhysRevA.76.044103}
\begin{equation}
\delta x_{w}\propto gRe\left[\langle A\rangle_{w}\right],
\end{equation}
and
\begin{equation}
\delta p_{w}=2gIm\left[\langle A\rangle_{w}\right]Var\left(P\right),
\end{equation}
respectively, where $Var\left(P\right)=\langle\Phi\vert P^{2}\vert\Phi\rangle-\langle\Phi\vert P\vert\Phi\rangle^{2}$
is the variance of the momentum operator $\hat{P}$ on the initial
pointer state, and $\langle A\rangle_{w}$ is the weak value, which
is defined as
\begin{align}
\langle A\rangle_{w} & =\frac{\langle\psi_{i}\vert\hat{A}\vert\psi_{f}\rangle}{\langle\psi_{i}\vert\psi_{f}\rangle}=\frac{\sum_{j}\sum_{k}\alpha_{j}\beta_{k}^{\ast}\langle a_{k}\vert\hat{A}\vert a_{j}\rangle}{\sum_{j}\sum_{k}\alpha_{j}\beta_{k}^{\ast}\langle a_{k}\vert a_{j}\rangle}\nonumber \\
 & =\frac{\sum_{j}a_{j}\alpha_{j}\beta_{j}^{\ast}}{\sum_{j}\alpha_{j}\beta_{j}^{\ast}}\,,\label{eq:13}
\end{align}
where, $Re[c]$ and $Im[c]$ denote the real and the imaginary parts
of the complex number $c$, respectively. In
general, the conditional expectation value $\langle A\rangle_{c}$
and the weak value $\langle A\rangle_{w}$ are different {[}see
Eqs. (\ref{eq:6}) and  (\ref{eq:13}) {]}, and thus correspond
to different measurement strengths \citep{Ban140}. In other words,
the (conditional) expectation value of system observable is related
to the (postselected) strong measurement, while the weak value is
caused by the postselected weak measurement. In can be easily seen
that if $\beta_{i}=\alpha_{i},$ the conditional
expectation value and the weak value reducedto the expectation
value given in Eq. (\ref{eq:6-1}). As shown above, these values can
be obtained from the position shifts of the pointer after the measurement.

\section{General Approach to the weak-to-strong measurement transition }

\begin{figure}
$\ \ \ \ \ $\includegraphics[width=8cm]{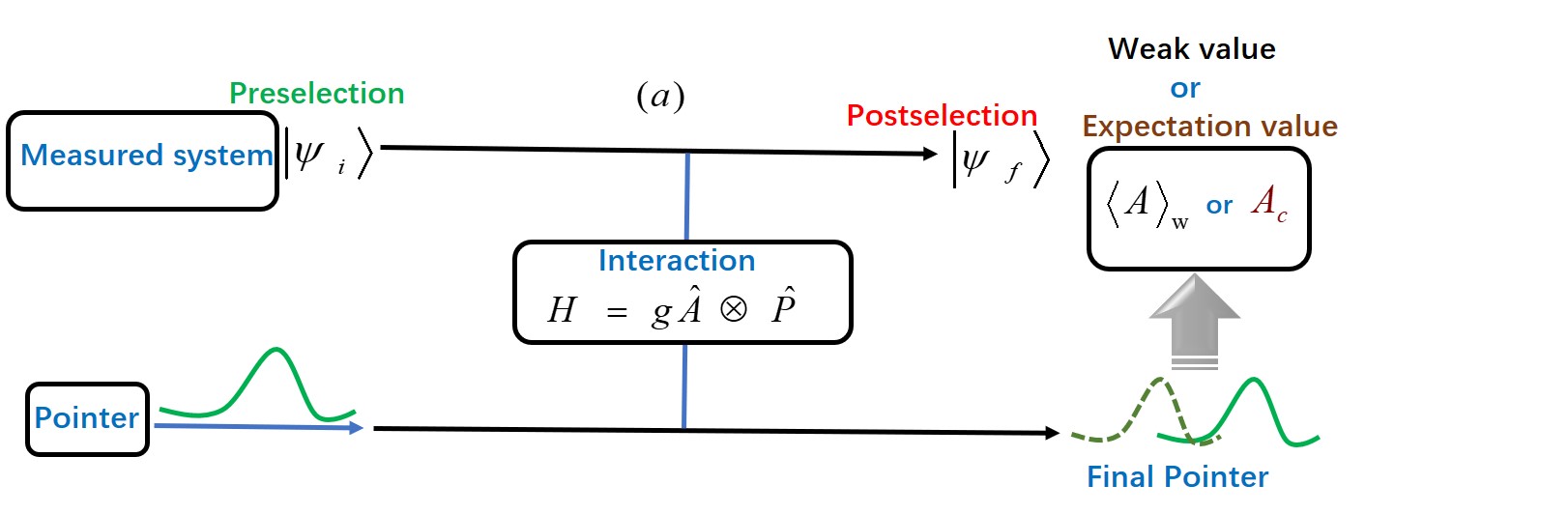}

\vspace{0.2cm}

\includegraphics[width=8cm]{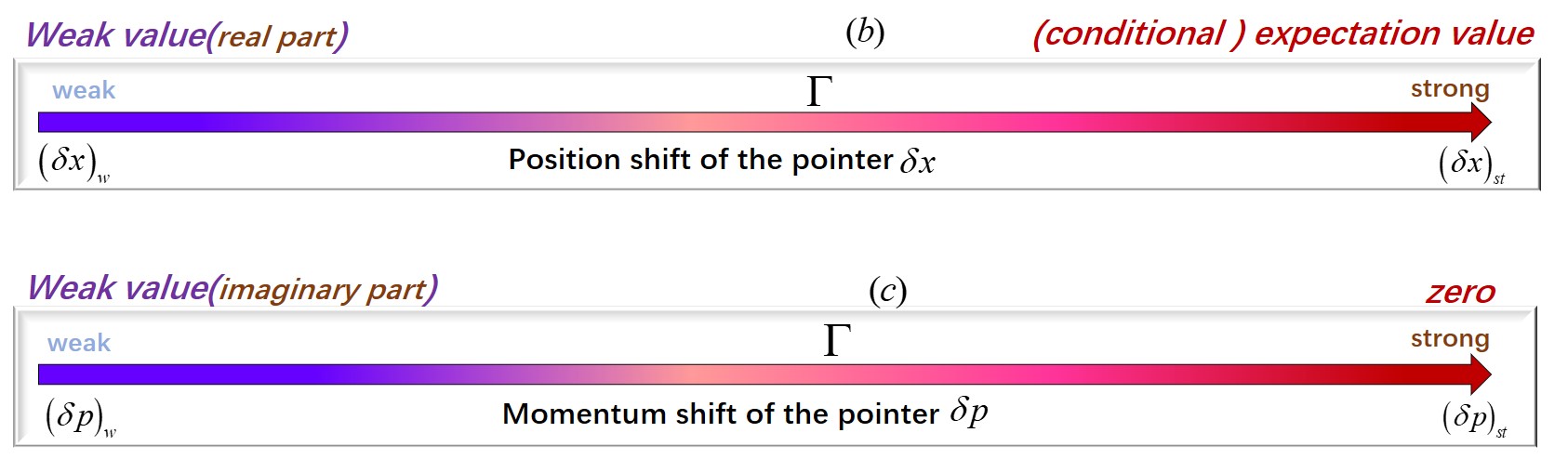}

\caption{\label{fig:0}Schematic diagram of our scheme to realize
weak-to-strong measurement transition . Panel (a) shows the postselected von Neumann measurement model; (b) and (c) show the relationships between
the shifts of the pointer and the values of the associated system
 observable. }
\end{figure}

In some recent studies, the weak-to-strong measurement transition has been investigated theoretically and experimentally by using some specific
pointer state \citep{Nature2020,PhysRevA.103.052215}.
However, a general approach to this problem has not been presented
yet. In this section, we provide a general approach based on the
measurement scheme illustrated in Fig. \ref{fig:0}.
We know that the Fock number states are the eigenstates
of the number operator $\hat{n}=\hat{a}^{\dagger}\hat{a}$, i.e.,
$\hat{a}^{\dagger}\hat{a}\vert n\rangle=n\vert n\rangle$. One can
expand any state in the basis of $\vert n\rangle$ since the
number operator is Hermitian. In other words, any state $\vert\phi\rangle$
can be written as:
\begin{equation}
\vert\phi\rangle=\sum_{n}c_{n}\vert n\rangle,
\end{equation}
where the expansion coefficients $c_{n}$ are determined by $c_{n}=\langle n\vert\phi\rangle$.
Let us assume that the pointer and the measured
system are prepared in the states $\vert\phi\rangle$ and $\vert\psi_{i}\rangle,$
respectively. Then the total initial state can be written
as $\vert\phi\rangle\otimes\vert\psi_{i}\rangle$ and the evolution
of the total system is generated by the interaction
Hamiltonian given in Eq. (\ref{eq:H-1}) as
\begin{align}
\vert\Psi_{evol}\rangle & =\exp\left(-i\int_{0}^{t}H_{int}(\tau)d\tau\right)\vert\psi_{i}\rangle\otimes\vert\phi\rangle\nonumber \\
 & =\sum_{i}\alpha_{i}D\left(\frac{\Gamma}{2}a_{i}\right)\vert a_{i}\rangle\otimes\vert\phi\rangle\nonumber \\
 & =\sum_{n,i}\alpha_{i}c_{n}\vert a_{i}\rangle\otimes\left|\frac{\Gamma}{2}a_{i},n\right\rangle ,\label{eq:15}
\end{align}
where
\begin{equation}
D\left(\frac{\Gamma}{2}a_{i}\right)=\exp\left(\frac{g}{2\sigma}a_{i}(a^{\dagger}-a)\right),\ \ \ \Gamma=\frac{g}{\sigma}
\end{equation}
is the displacement operator, and
\begin{equation}
\left|\frac{\Gamma}{2}a_{i},n\right\rangle =D\left(\frac{\Gamma}{2}a_{i}\right)\vert n\rangle
\end{equation}
is usually referred to as displaced Fock states. In the derivation
of Eq. (\ref{eq:15}), we used the expression of the pointer momentum operator
 $\hat{P}$ in terms
of the annihilation and creation operators $\hat{a}$ and $\hat{a}^{\dagger}$as
\begin{equation}
\hat{P}=\frac{i}{2\sigma}\left(\hat{a}^{\dagger}-\hat{a}\right).
\end{equation}
 The parameter $\Gamma=g/\sigma$ characterizes the coupling strength
between the measured system and the pointer. If $\Gamma\gg1$ ($\Gamma\ll1$),
the measurement is in the strong (weak) measurement regime. After
a postselection onto the state $\vert\psi_{f}\rangle$
the normalized final state of the pointer reads as follows
\begin{equation}
\vert\varTheta\rangle=\mathcal{N}\sum_{i}\sum_{n}\alpha_{i}\beta_{i}^{\ast}c_{n}\left|\frac{\Gamma}{2}a_{i},n\right\rangle ,\label{eq:19}
\end{equation}
where $\mathcal{N}$ is a normalization coefficient given by
\begin{align}
\mathcal{N}^{-2} & =\sum_{i,j}\sum_{n,m}c_{m}^{\ast}c_{m}\alpha_{j}^{\ast}\beta_{j}\alpha_{i}\beta_{i}^{\ast}L_{n}\left(\frac{\Gamma^{2}}{4}\vert a_{i}-a_{j}\vert^{2}\right)\times\nonumber \\
 & \exp\left[-\frac{\Gamma^{2}}{8}\vert a_{i}-a_{j}\vert^{2}\right]\nonumber \\
 & =\sum_{i}\vert\alpha_{i}\beta_{i}^{\ast}\vert^{2}+\sum_{n}\sum_{i\neq j}\alpha_{j}^{\ast}\beta_{j}\alpha_{i}\beta_{i}^{\ast}\vert c_{n}\vert^{2}L_{n}\left(\frac{\Gamma^{2}}{4}\vert a_{i}-a_{j}\vert^{2}\right)\times\nonumber \\
 & \exp\left[-\frac{\Gamma^{2}}{8}\vert a_{i}-a_{j}\vert^{2}\right],\label{eq:20}
\end{align}
where $L_{n}(x)$ denote the n-th Laguerre polynomials. Eq. (\ref{eq:19})
contains the information about the system observable $A$ that we want to
measure, and holds for all Fock-state-based pointer states. In the
next subsection, we derive the general formulas of pointer shifts
and discuss the weak-to-strong measurement transition.

\subsection{Position shift}

Here, we give the general expression of the position shift of the
Fock-state-based pointer states after the postselected von Neumann
measurement. The position operator $\hat{X}$ of the pointer can be
written in term of annihilation and creation operators $\hat{a}$
and $\hat{a}^{\dagger}$as

\begin{equation}
\hat{X}=\sigma\left(\hat{a}^{\dagger}+\hat{a}\right).
\end{equation}
 Since the interaction causes the shifts of the pointer, we can read
the system information by comparing the pointer state
before and after the measurement. For our scheme, the position shift
of the pointer after the postselected von Neumann measurement reads
as
\begin{align}
\delta x & =\langle\varTheta\vert\hat{X}\vert\varTheta\rangle-\langle\phi\vert\hat{X}\vert\phi\rangle\nonumber \\
 & =\mathcal{\vert N}\vert^{2}\sigma\sum_{i,j}\sum_{n,m}c_{m}^{\ast}c_{n}\alpha_{j}^{\ast}\beta_{j}\alpha_{i}\beta_{i}^{\ast}\nonumber \\
 & \text{\ensuremath{\times}}\{\sqrt{n+1}\langle m\vert D\left[\frac{\Gamma}{2}\left(a_{i}-a_{j}\right)\right]\vert n+1\rangle\nonumber \\
 & +\sqrt{n}\langle m\vert D\left[\frac{\Gamma}{2}\left(a_{i}-a_{j}\right)\right]\vert n-1\rangle+\nonumber \\
 & \Gamma a_{i}\langle m\vert D\left[\frac{\Gamma}{2}\left(a_{i}-a_{j}\right)\right]\vert n\rangle\}-\langle\phi\vert\hat{X}\vert\phi\rangle,\label{eq:22}
\end{align}
where
\begin{equation}
\langle\phi\vert\hat{X}\vert\phi\rangle=2\sigma\sum_{n}\sqrt{n+1}Re\left[c_{n}c_{n+1}^{\ast}\right]\label{eq:23}
\end{equation}
is the initial average value of the position operator on the initial
pointer state $\vert\phi\rangle$. The explicit forms of the matrix
elements of displacement operator in the above formula are given by \citep{PhysRevA.41.2645}
\begin{align}
 & \langle m\vert D\left(\alpha\right)\vert n\rangle\nonumber \\
 & =e^{-\frac{\vert\alpha\vert^{2}}{2}}\begin{cases}
\sqrt{\frac{n!}{m!}}L_{n}^{m-n}\left(\vert\alpha\vert^{2}\right)\left(\alpha\right)^{m-n}, & m\ge n\\
\sqrt{\frac{m!}{n!}}L_{m}^{n-m}\left(\vert\alpha\vert^{2}\right)\left(\alpha\right)^{n-m}. & m\leq n
\end{cases}\label{eq:26-1}
\end{align}
where the generalized Laguerre polynomials are defined as
\begin{equation}
L_{n}^{(\eta)}(x)=\sum_{k=0}^{n}\left(\begin{array}{c}
n+\eta\\
n-k
\end{array}\right)\frac{(-1)^{k}}{k!}x^{k}.\label{eq:27-1}
\end{equation}
Eq. (\ref{eq:22}) is the general expression of the position shift
of the Fock- state-based pointer measurement schemes, and it holds
in all measurement regimes.

If one wants to know the position shift of the pointer in the postselected
strong measurement regime, then one can consider larger values
of the coupling  $\Gamma$. In order to do that,
a limit $\Gamma\rightarrow\infty$ should be taken to Eq. (\ref{eq:22}
) which gives

\begin{align}
\left(\delta x\right)_{st} & =\lim_{\Gamma\rightarrow\infty}\delta x\nonumber \\
= & \frac{\text{\ensuremath{\sigma}}}{\sum_{i}\vert\alpha_{i}\beta_{i}^{\ast}\vert^{2}}\{\sum_{i}\sum_{n,m}c_{m}^{\ast}c_{n}\vert\alpha_{i}\beta_{i}^{\ast}\vert^{2}\times\nonumber \\
 & \left(\sqrt{n+1}\delta_{m,n+1}+\sqrt{n}\delta_{m,n-1}+\Gamma a_{i}\delta_{m,n}\right)\}\nonumber \\
 & -\langle\phi\vert\hat{X}\vert\phi\rangle\nonumber \\
 & =g\frac{\sum_{i}a_{i}\vert\alpha_{i}\beta_{i}^{\ast}\vert^{2}}{\sum_{i}\vert\alpha_{i}\beta_{i}^{\ast}\vert^{2}}=g\langle A\rangle_{c}.\label{eq:24}
\end{align}
In the above expression, $\delta_{m,n}$ represents the Kronecker
delta function
\begin{equation}
\delta_{m,n}=\begin{cases}
1, & m=n\\
0, & m\neq n
\end{cases}\label{eq:26-2}
\end{equation}
Eq. (\ref{eq:24}) is the position shift formula for the postselected
strong von Neumann measurement introduced in Sec. \ref{sec:2}.
On the other hand, if we are interested in the postselected weak measurement,
$\Gamma$ should be set vanishingly  small. In this  case,
the position shift of the pointer reads as follows
\begin{align}
\left(\delta x\right)_{w} & =\lim_{\Gamma\rightarrow0}\delta x\nonumber \\
 & =gtRe\left[\langle A\rangle_{w}\right]+2gtIm\left[\langle A\rangle_{w}\right]\times\nonumber \\
 & \{Im\left[\sum_{n}\sqrt{(n+1)(n+2)}c_{n}^{\ast}c_{n+2}\right]\nonumber \\
 & -Im\left[\sum_{n}\sqrt{n+1}c_{n}^{\ast}c_{n+1}\right]^{2}\}\nonumber \\
 & =gRe\left[\langle A\rangle_{w}\right]+2gIm\left[\langle A\rangle_{w}\right]Im\left[\langle a^{2}\rangle-\langle a\rangle^{2}\right].\label{eq:25}
\end{align}
Eq. (\ref{eq:25}) is the concise form of the expression
presented in \citep{PhysRevA.85.012107}. It gives the position shift
of the pointer in posteselectd weak measurement, and holds for any
pointer state that can be written in Fock state basis. In the above position
shift formula, $\langle a\rangle$ ($\langle a^{2}\rangle$) denotes
the expectation value of the annihilation operator $a$ (squared annihilation
operator $a$) in the initial pointer state $\vert\phi\rangle$.

Let us now exploit Eq. (\ref{eq:25}) to investigate the transition for some specific pointer state, which have been investigated in recent studies {cite those studies here?}
\textsl{(1) Coherent state}. If we assume that the pointer
is initially prepared in a coherent state, i.e., $\vert\phi\rangle=\vert\alpha\rangle=\sum_{n}c_{n}\vert n\rangle$
with $c_{n}=e^{-\frac{\vert\alpha\vert^{2}}{2}}\frac{\alpha^{n}}{\sqrt{n!}}$,
then
\begin{equation}
\langle a^{2}\rangle_{ch}=\langle\alpha\vert a^{2}\vert\alpha\rangle=\alpha^{2},\ \ \ \langle\alpha\vert a\vert\text{\ensuremath{\alpha\rangle=\alpha} .}\label{eq:26}
\end{equation}
In this case the position shift, Eq. (\ref{eq:25}), is given by
\begin{equation}
\left(\delta x\right)_{w,coh}=gRe\left[\langle A\rangle_{w}\right].\label{eq:27}
\end{equation}
This result coincides with that presented in Ref. \citep{PhysRevA.92.022109,PhysRevA.105.022210}.

\textsl{(2)} \textsl{Coherent squeezed state.} Assuming the pointer initially prepared in a coherent squeezed state, we recover the results of
Ref. \citep{PhysRevA.103.052215} by using Eq. (\ref{eq:25}).
Squeezed coherent states are defined as $\vert\phi\rangle=\vert\alpha,\xi\rangle=D(\alpha)S(\xi)\vert0\rangle$
where
\begin{equation}
D(\alpha)=\exp\left(\alpha a^{\dagger}-\alpha^{\ast}a\right),S(\xi)=\exp\left(\frac{1}{2}\xi^{\ast}a^{2}-\frac{1}{2}\xi a^{\dagger2}\right).\label{eq:28}
\end{equation}
Here, $D(\alpha)$ and $S(\xi)$ are the displacement and squeezed
operator, respectively, with $\alpha=\vert\alpha\vert e^{i\phi_{\alpha}}$
and $\xi=re^{i\phi_{\xi}}$. This state also can be written in terms
of Fock state basis as \citep{GerryIntroductory}
\begin{equation}
\vert\alpha,\xi\rangle=\sum_{n}c_{n}\vert n\rangle,
\end{equation}
where
\begin{align}
c_{n} & =\frac{1}{\sqrt{\cosh r}}\exp\left[-\frac{\vert\alpha\vert^{2}}{2}-\frac{1}{2}\alpha^{\ast2}e^{i\phi_{\xi}}\tanh r\right]\nonumber \\
 & \times\sum_{n=0}^{\infty}\frac{\left[\frac{1}{2}e^{i\phi_{\xi}}\tanh r\right]^{n/2}}{\sqrt{n!}}H_{n}\left[\gamma\left(e^{i\phi_{\xi}}\sinh(2r)\right)^{-\frac{1}{2}}\right]
\end{align}
with $\gamma=\alpha\cosh r+\alpha^{\ast}e^{i\phi_{\xi}}\sinh r$,
and $H_{n}(x)$ being the the order $n$ Hermite polynomial. The expectation
value of $\langle a\rangle$ and $\langle a^{2}\rangle$ under the
state $\vert\alpha,\xi\rangle$ are given as
\begin{equation}
\langle a\rangle_{sc}=\alpha,\ \ \ \langle a^{2}\rangle_{sc}=\alpha^{2}-e^{i\phi_{\xi}}\sinh r\cosh r.\label{eq:30}
\end{equation}
In this case, after the postselected weak measurement, the position
shift of the pointer reads as
\begin{align}
\left(\delta x\right)_{w,sc} & =gRe\left[\langle A\rangle_{w}\right]-gIm\left[\langle A\rangle_{w}\right]\sinh2r\sin\phi_{\xi}.\label{eq:31}
\end{align}
This is the same result derived in Ref. \citep{PhysRevA.103.052215}
for squeezed coherent pointer.

\textsl{(3) Single-photon-added coherent state(SPAC).} The SPAC state
is defined as \citep{PhysRevA.43.492}

\begin{equation}
\vert1,\alpha\rangle=\sum_{n}c_{n}\vert n+1\rangle\label{eq:36}
\end{equation}
where
\begin{equation}
c_{n}=\frac{e^{-\frac{\vert\alpha\vert^{2}}{2}}}{\sqrt{1+\vert\alpha\vert^{2}}}\sqrt{\frac{n+1}{n!}}.
\end{equation}
The expectation values $\langle a\rangle_{spac}$ and $\langle a^{2}\rangle_{spac}$
for the state $\text{\ensuremath{\vert1,\alpha\rangle}}$ are given by
\begin{equation}
\langle a\rangle_{spac}=\frac{\alpha(2+\vert\alpha\vert^{2})}{1+\vert\alpha\vert^{2}},
\end{equation}
and
\begin{equation}
\langle a^{2}\rangle_{spac}=\frac{\alpha^{2}(3+\vert\alpha\vert^{2})}{1+\vert\alpha\vert^{2}},\ \ \ \
\end{equation}
respectively. Thus, if we take the pointer initially prepared in a SPAC state, its position
shift after the postselected weak measurement would be shifted by
\begin{equation}
\left(\delta x\right)_{w,spac}=gRe\left[\langle A\rangle_{w}\right]-2g\frac{\vert\alpha\vert^{2}\sin2\phi_{\alpha}}{\left(1+\vert\alpha\vert^{2}\right)^{2}}Im\left[\langle A\rangle_{w}\right].
\end{equation}
This is the same result obtained in Ref. \citep{PhysRevA.105.022210,Yusuf2022}.
From the above examples it can be confirmed that the Eq. ( \ref{eq:25})
can be used for any pointer states.

\subsection{Momentum shift}
The shifts in the momentum of the pointer may be evaluated in a similar way.
After a postselected von Neumann measurement we have
\begin{align}
\text{\ensuremath{\delta p} } & =\langle\varTheta\vert\hat{P}\vert\varTheta\rangle-\langle\phi\vert\hat{P}\vert\phi\rangle\nonumber \\
 & =\frac{i}{2\sigma}\mathcal{\vert N}\vert^{2}\sum_{i,j}\sum_{n,m}c_{m}^{\ast}c_{n}\alpha_{j}^{\ast}\beta_{j}\alpha_{i}\beta_{i}^{\ast}\nonumber \\
 & \times\{\sqrt{n+1}\langle m\vert D\left[\frac{\Gamma}{2}\left(a_{i}-a_{j}\right)\right]\vert n+1\rangle\nonumber \\
 & -\sqrt{n}\langle m\vert D\left[\frac{\Gamma}{2}\left(a_{i}-a_{j}\right)\right]\vert n-1\rangle\}-\langle\phi\vert\hat{P}\vert\phi\rangle\,,\label{eq:42}
\end{align}
where
\begin{equation}
\langle\phi\vert\hat{P}\vert\phi\rangle=\frac{1}{\sigma}Im\left[\sum_{n}\sqrt{n+1}c_{n+1}^{\ast}c_{n}\right]
\end{equation}
is the expectation value of the momentum operator in the initial
pointer state $\vert\phi\rangle.$ No approximation has been used to derive
 Eq. ( \ref{eq:42}), and thus it provides the
momentum shift of the pointer in all measurement regimes.
Again, the two extremes give the momentum shifts corresponding to
the strong and weak measurement, respectively: (i) For a very strong
coupling ($\Gamma\gg1)$, corresponding to the strong measurement regime,
the momentum shift becomes
\begin{align}
\left(\delta p\right)_{st} & =\lim_{\Gamma\rightarrow\infty}\delta p\nonumber \\
 & =\frac{i}{2\sigma\sum_{j}\vert\alpha_{i}\beta_{i}\vert^{2}}\sum_{i,j}\sum_{n,m}c_{m}^{\ast}c_{n}\alpha_{j}^{\ast}\beta_{j}\alpha_{i}\beta_{i}^{\ast}\nonumber \\
 & \times\{\sqrt{n+1}\langle m\vert D\left[\frac{\Gamma}{2}\left(a_{i}-a_{j}\right)\right]\vert n+1\rangle\nonumber \\
 & -\sqrt{n}\langle m\vert D\left[\frac{\Gamma}{2}\left(a_{i}-a_{j}\right)\right]\vert n-1\rangle\}-\langle\phi\vert\hat{P}\vert\phi\rangle\nonumber \\
 & =\frac{i}{2\sigma}\sum_{n,m}c_{m}^{\ast}c_{n}\{\sqrt{n+1}\langle m\vert n+1\rangle-\sqrt{n}\langle m\vert n-1\rangle\}\nonumber \\
 & -\frac{1}{\sigma}Im\left[\sum_{n}\sqrt{n+1}c_{n+1}^{\ast}c_{n}\right]=0.\label{eq:43}
\end{align}
 (ii) On the contrary, in the weak measurement regime ($\Gamma\ll1$),
we get
\begin{align}
\left(\delta p\right)_{w} & =\lim_{\Gamma\rightarrow0}\delta p\nonumber \\
 & =\frac{i}{2\text{\ensuremath{\sigma\sum_{i,j}\alpha_{j}^{\ast}\beta_{j}\alpha_{i}\beta_{i}^{\ast}}}}\sum_{i,j}\sum_{n,m}c_{m}^{\ast}c_{n}\alpha_{j}^{\ast}\beta_{j}\alpha_{i}\beta_{i}^{\ast}\nonumber \\
 & \times\{\sqrt{n+1}\langle m\vert D\left[\frac{\Gamma}{2}\left(a_{i}-a_{j}\right)\right]\vert n+1\rangle\nonumber \\
 & -\sqrt{n}\langle m\vert D\left[\frac{\Gamma}{2}\left(a_{i}-a_{j}\right)\right]\vert n-1\rangle\}-\langle\phi\vert\hat{P}\vert\phi\rangle\nonumber \\
 & =2gIm\left[\langle A\rangle_{w}\right]\left(\langle\hat{P}^{2}\rangle-\langle\hat{P}\rangle^{2}\right)\nonumber \\
 & =2gIm\left[\langle A\rangle_{w}\right]Var(P).\label{eq:47}
\end{align}
 where $Var(P)$ is the variance of the momentum operator $\hat{P}$
in the initial pointer state $\vert\phi\rangle$. The final results
of Eq. (\ref{eq:43}) and Eq. (\ref{eq:47}) agree with those contained in previous works \citep{PhysRevA.103.052215,PhysRevA.76.044103,Yusuf2022},
and holds for any pointer states.

\section{\label{sec:3}Example: coherent state pointer case}

\subsection{Measurement transition}
As an example of application of our general method to described
the weak-to-strong measurement transition, we here consider a pointer state initially prepared in a coherent state. We assume
that the interaction Hamiltonian between the measured system and the
pointer has the form (\ref{eq:H-1}), and the system
observable $\hat{A}$ is the Pauli-$x$ operator, i.e., $\hat{A}=\hat{\sigma}_{x}=\vert+\rangle\langle+\vert+\vert-\rangle\langle-\vert$.
Here, $\vert\pm\rangle\equiv\left(\vert\uparrow\rangle\pm\vert\downarrow\rangle\right)/\sqrt{2}$
are the eigenstates of $\hat{\sigma}_{x}$ with eigenvalues $\pm1$,
respectively. If we assume that the measured system is initially
prepared in the state $\vert\psi_{ini}\rangle=\vert\downarrow\rangle$, and
pointer is prepared in coherent state $\vert\phi(x)\rangle=\vert\beta\rangle$
with $\beta=re^{i\varphi}$, then the initial total state of the system
can be written as $\vert\downarrow\rangle\otimes\vert\beta\rangle$.
Given the unitary operator $U=e^{-ig\hat{\sigma}_{x}\otimes\hat{P}}$,
the total system state given in Eq. (\ref{eq:15}) evolves as
\begin{align}
\vert\Omega\rangle & =\frac{1}{\sqrt{2}}\left[\vert+\rangle D\left(\frac{\Gamma}{2}\right)\vert\beta\rangle-\vert-\rangle D^{\dagger}\left(\frac{\Gamma}{2}\right)\vert\beta\rangle\right],\label{eq:4}
\end{align}
with $\Gamma=\frac{g}{\sigma}.$ If we take a postselection onto the
state $\vert\psi_{f}\rangle=\cos\theta\vert\uparrow\rangle-\sin\theta\vert\downarrow\rangle$,
then the final state of the pointer [see Eq. (\ref{eq:19}]
reduces to

\begin{align}
\!\!\!\!\! & \vert\Psi^{\prime\prime}\rangle=\nonumber \\
\!\!\!\!\! & \frac{\sin\left(\frac{\pi}{4}-\theta\right)e^{-i\frac{\Gamma}{2}r\sin\varphi}\vert\beta+\frac{\Gamma}{2}\rangle-\cos\left(\frac{\pi}{4}-\theta\right)e^{i\frac{\Gamma}{2}r\sin\varphi}\vert\beta-\frac{\Gamma}{2}\rangle}{\sqrt{1-\cos\left(2\theta\right)\cos\left(2\Gamma r\sin\varphi\right)e^{-\frac{1}{2}\Gamma^{2}}}}.\label{eq:49}
\end{align}
As we can see, this is a superposition of two coherent states
with different coherent amplitudes. In the $x$-representation the state
$\vert\Psi^{\prime}\rangle$ can be expressed as

\begin{align}
\Psi^{\prime\prime}(x) & =\langle x\vert\Psi^{\prime\prime}\rangle=\nonumber \\
\!\!\!\!\!\!\!\!\!\!\!\!\!\!\! & \frac{\sin\left(\frac{\pi}{4}-\!\!\theta\right)e^{-i\frac{\Gamma}{2}r\sin\varphi}\psi_{+}(x)\!-\!\cos\left(\frac{\pi}{4}-\!\!\beta\right)e^{i\frac{\Gamma}{2}r\sin\varphi}\psi_{-}(x)}{\sqrt{1-\cos\left(2\theta\right)\cos\left(2\Gamma r\sin\varphi\right)e^{-\frac{1}{2}\Gamma^{2}}}},\label{eq:50-2}
\end{align}
where

\begin{align}
\psi_{\pm}(x) & =\langle x\vert\beta\pm\frac{\Gamma}{2}\rangle\nonumber \\
 & =\left(\frac{1}{2\pi\sigma^{2}}\right)^{\frac{1}{4}}e^{-\vert\beta\pm\frac{\Gamma}{2}\vert^{2}}\nonumber \\
 & \times\exp\left(\frac{x^{2}}{4\sigma^{2}}\right)\exp\left[-\frac{\left(x-\sigma(\beta\pm\frac{\Gamma}{2})\right)^{2}}{2\sigma^{2}}\right]\label{eq:50-1}
\end{align}
are Gaussian wave packets  with central position shifts $\pm\frac{\Gamma}{2}$ compared to the initial Gaussian caused
by the interaction with the measured system. On the other hand, for
our pre- and postselected states $\vert\downarrow\rangle$ and $\cos\theta\vert\uparrow\rangle-\sin\theta\vert\downarrow\rangle$
, the corresponding conditional expectation value and weak value defined
in Eq. (\ref{eq:6}) and Eq. (\ref{eq:13}) can be written as
\begin{equation}
\langle\sigma_{x}\rangle_{c}=-\sin(2\theta),
\end{equation}
and
\begin{equation}
\langle\sigma_{x}\rangle_{w}=-\cot(\theta),
\end{equation}
respectively. As we show in the following, the conditional expectation
value $\langle\sigma_{x}\rangle_{c}$ and weak value $\langle\sigma_{x}\rangle_{w}$
are directly related to  the pointer shifts of strong postselected measurement
and postselected weak measurement, respectively.

\begin{figure}
\includegraphics[width=8cm]{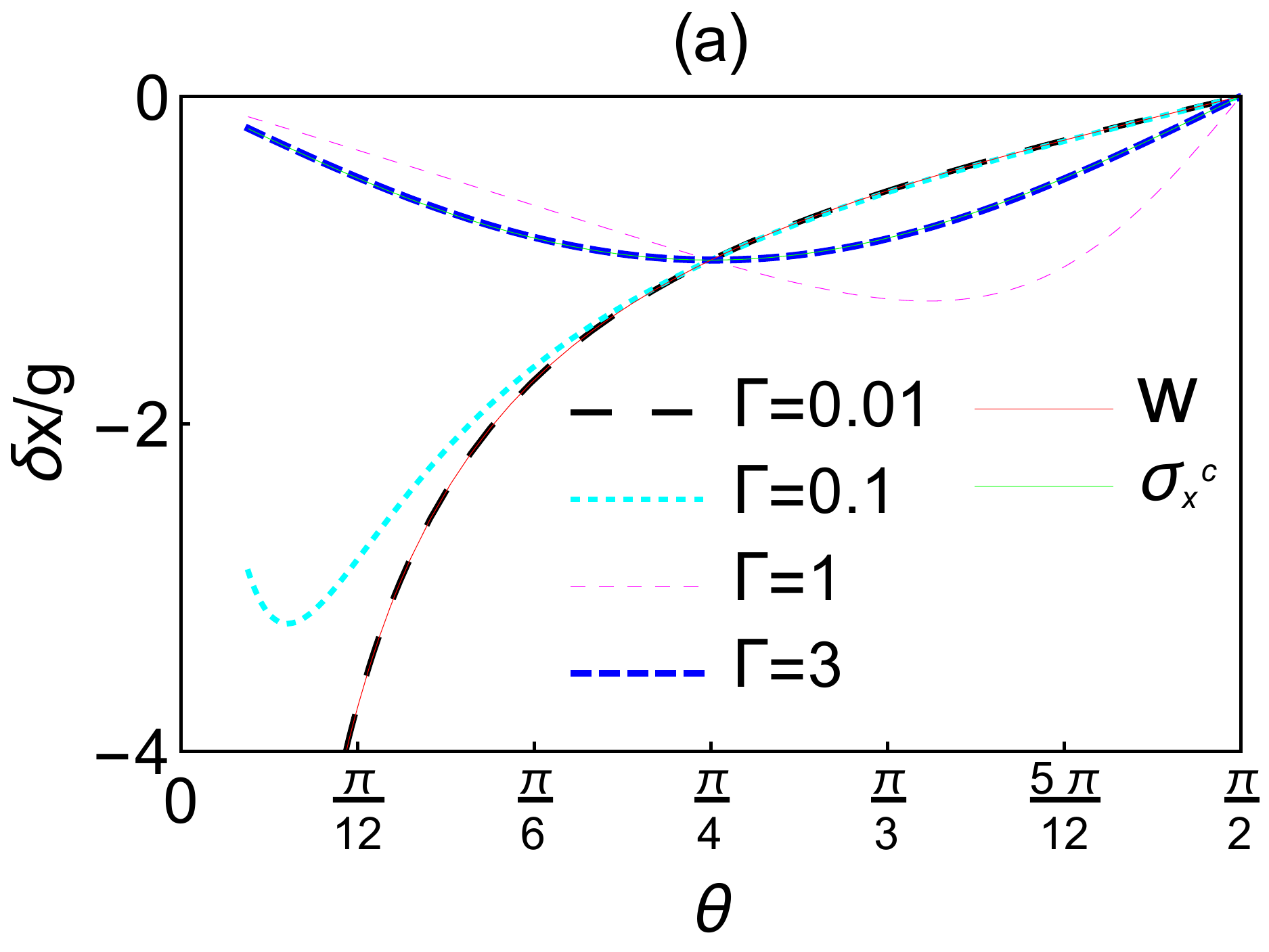}

\includegraphics[width=8cm]{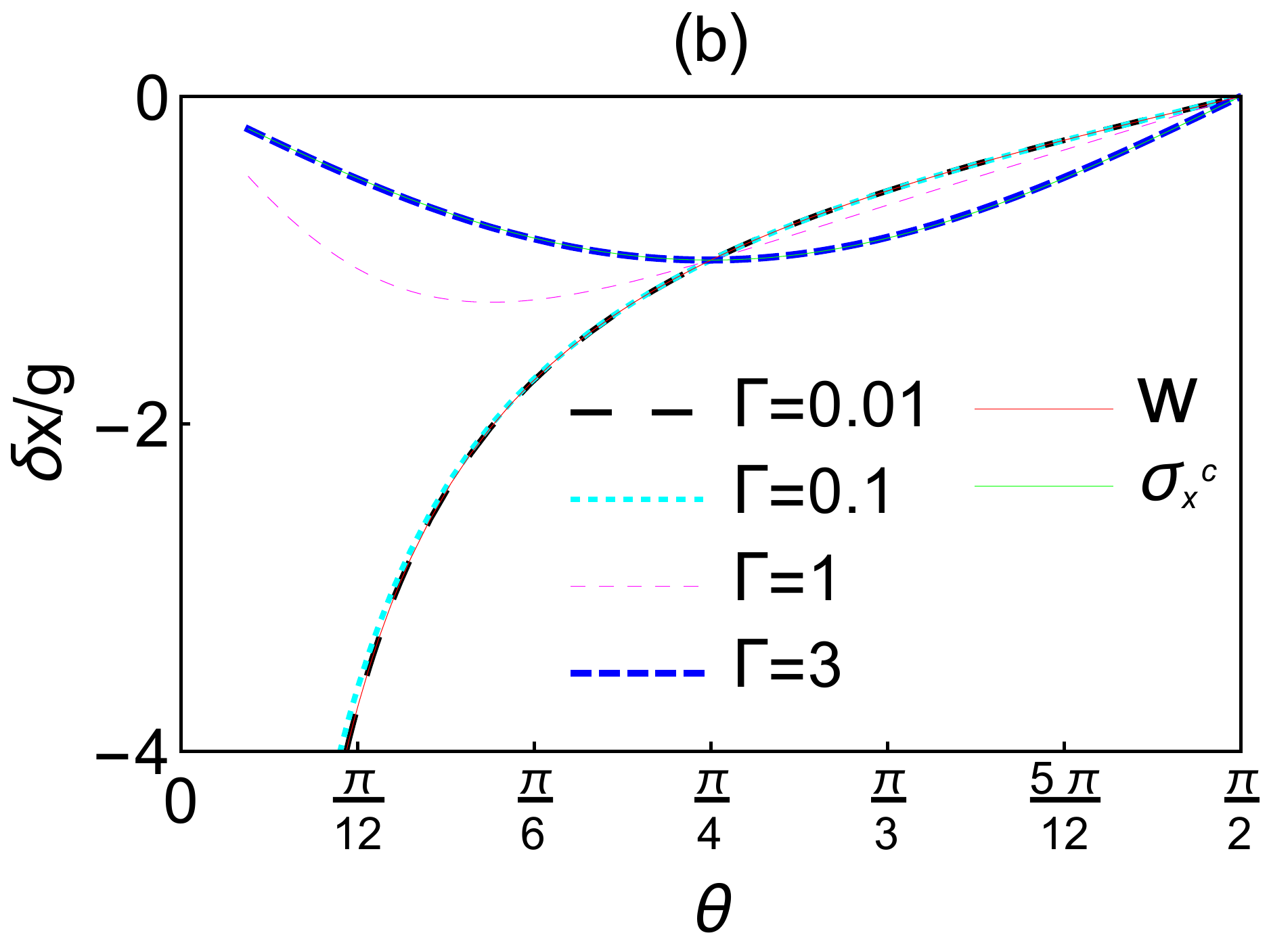}

\caption{\label{fig:1}Weak-to-strong measurement transition, as characterized
by position shift (see the Eq.(\ref{eq:52-1})) of the coherent pointer
state. The $\delta x_{coh}$ is shown as function of system postselected
state parameter $\theta$ for different values of the coupling strengths
$\Gamma$: (a) $r=3$ and (b) $r=0$. Here, we take $\varphi=\frac{\pi}{6}$. }
\end{figure}

\textsl{Position shift.} We can check the weak-to-strong measurement
transition by investigating the position and momentum shifts of the
pointer after measurement. As mentioned in Sec. \ref{sec:2},
the the parameter $\Gamma$ quantifies the measurement
strength and it can takes all values, thus realizing a weak-to-strong transition.
The general expression of the position shift after postselected von
Neumann measurement for coherent pointer state, can be calculated
by using Eq. (\ref{eq:50-2}). We hav

\begin{align}
\delta x_{coh} & =\langle x\rangle_{fi}-\langle x\rangle_{ini}\nonumber \\
\!\!\!\!\! & =\frac{-\text{\ensuremath{g}\ensuremath{\sin}\ensuremath{(2\theta)}}+\frac{1}{2}gr\cos\varphi\cos\left(2\theta\right)\cos\left(2\Gamma r\sin\varphi\right)e^{-\frac{1}{2}\Gamma^{2}}}{1-\cos\left(2\theta\right)\cos\left(2\Gamma r\sin\varphi\right)e^{-\frac{1}{2}\Gamma^{2}}}.\label{eq:52-1}
\end{align}
It is straightforward to see from this expression that the measurement
transition can be controlled by adjusting $\Gamma$ in the term
$e^{-\frac{1}{2}\Gamma^{2}}$. To get the position shift of the
coherent state pointer after a strong measurement, we take the limit
of Eq. (\ref{eq:52-1}) for $\Gamma$ goes to infinity, i.e.,
\begin{align}
\left(\delta x\right)_{strong} & =\lim_{\Gamma\rightarrow\infty}\delta x_{coh}=-g\text{\ensuremath{\sin}\ensuremath{(2\theta)}}=g\langle\sigma_{x}\rangle_{c}.\label{eq:53}
\end{align}
This result exactly matches with the Eq. (\ref{eq:24}). Taking the opposite
limit, i.e., $\Gamma\ll1$, we go to weak measurement regime and pointer shift position becomes
\begin{align}
\left(\delta x\right)_{weak} & =\lim_{\Gamma\rightarrow0}\delta x_{coh}=-g\cot\theta=g\langle\sigma_{x}\rangle_{w}.\label{eq:54}
\end{align}
 This result validates again Eq. ( \ref{eq:25}). In order
to explore the weak-to-strong measurement transition,  as characterized
by the position shift of the coherent pointer state, we have calculated $\delta x_{coh}$ as a function of postselected
system state parameter $\theta$ for various coupling strengths $\Gamma$. Results are shown in Fig. \ref{fig:1} and agree well with our theoretical predictions.
In panels (a) and (b) we report results for two values of $r$ and a fixed value of $\varphi$.

\begin{figure}
\includegraphics[width=8cm]{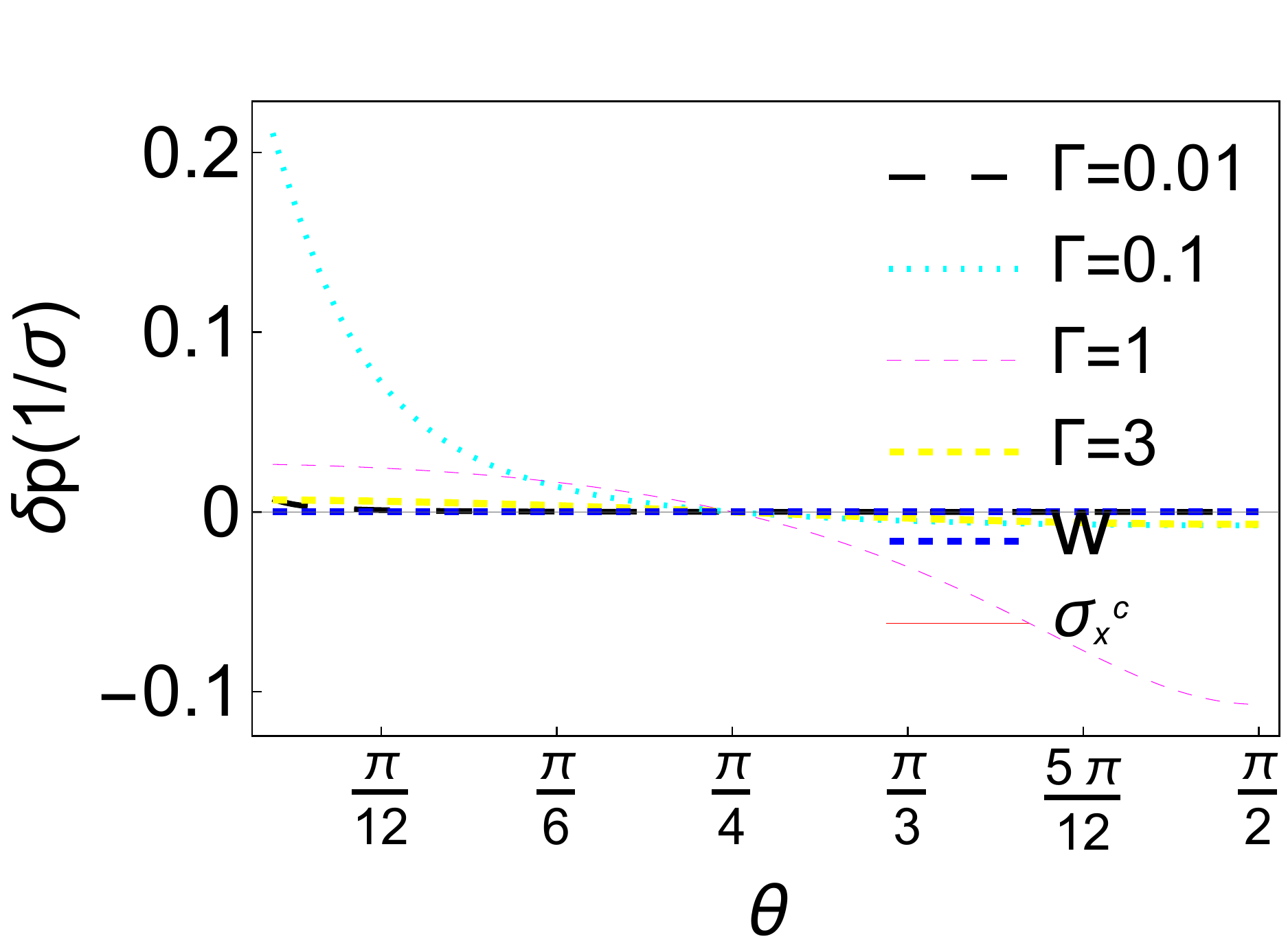}

\caption{\label{fig:2}Momentum shift of the coherent pointer state after postselected
von Neuamann measurement as function of system postselected state
parameter $\theta$ for various coupling strengths $\Gamma$. Here
we take $\varphi=\frac{\pi}{6}$ , $r=3$.}
\end{figure}

\textsl{Momentum shift}. The momentum shift of the coherent pointer
state also can be obtained by substituting the form of the state
into Eq. (\ref{eq:50-2}). Its explicit expression reads as follows

\begin{align}
\delta p_{coh} & =\langle\Psi^{\prime\prime}\vert\hat{P}\vert\Psi^{\prime\prime}\rangle-\langle\beta\vert\hat{P}\vert\beta\rangle\nonumber \\
 & =-\frac{1}{2\sigma\left(1-\cos\left(2\theta\right)\cos\left(2\Gamma r\sin\varphi\right)e^{-\frac{1}{2}\Gamma^{2}}\right)}\nonumber \\
 & \times\{2iIm(\beta^{\ast})-i\sin\left(2\Gamma r\sin\varphi\right)\cos(2\theta)e^{-\frac{1}{2}\Gamma^{2}}\Gamma\nonumber \\
 & \!\!-\!2iIm(\beta^{\ast})\cos(2\theta)\cos\left(2\Gamma r\sin\varphi\right)e^{-\frac{1}{2}\Gamma^{2}}\!\}\!+\!\frac{1}{2\sigma}2Im(\beta^{\ast}).\label{eq:50}
\end{align}
 It is obvious from the above expression that the factor $e^{-\frac{1}{2}\Gamma^{2}}$
can be used to control the measurement regimes in the same way as
the position shift. Similarly, we obtain
\begin{align}
\left(\delta p\right)_{weak} & =\lim_{\Gamma\rightarrow0}\delta p_{coh}=0\label{eq:51}
\end{align}
 and
\begin{align}
\left(\delta p\right)_{st} & =\lim_{\Gamma\rightarrow\infty}\delta p_{coh}=0,\label{eq:52}
\end{align}
respectively. These results well support our theoretical derivations
given in Sec. \ref{sec:2}. Notice that in this scheme the weak value
is real, whereas in general the momentum
shift of the pointer in postselected weak measurement is proportional
to the imaginary part of the weak value {[}see the Eq.(\ref{eq:47}){]}.
Thus, in our current case the momentum shift of the pointer for weak
measurement vanishes. In Fig. \ref{fig:2}, we show the the momentum
shift {[}see Eq. (\ref{eq:50}){]} as a function
of parameter $\theta$.

In Ref. \citep{Nature2020}, the authors have experimentally studied the
weak-to-strong measurement transition problem in $^{40}Ca^{+}$ trapped
ion system by considering a zero mean  Gaussian as a pointer. The
zero Gaussian profile corresponds to the ground state ($n=0$) of
the coherent pointer state. Thus, our discussion in this section can
be seen as an extension of the theoretical part of Ref. \citep{Nature2020}.

\subsection{The phase space function----- the Husimi-Kano Q function }

\begin{widetext}

\begin{figure}
\includegraphics[width=6cm]{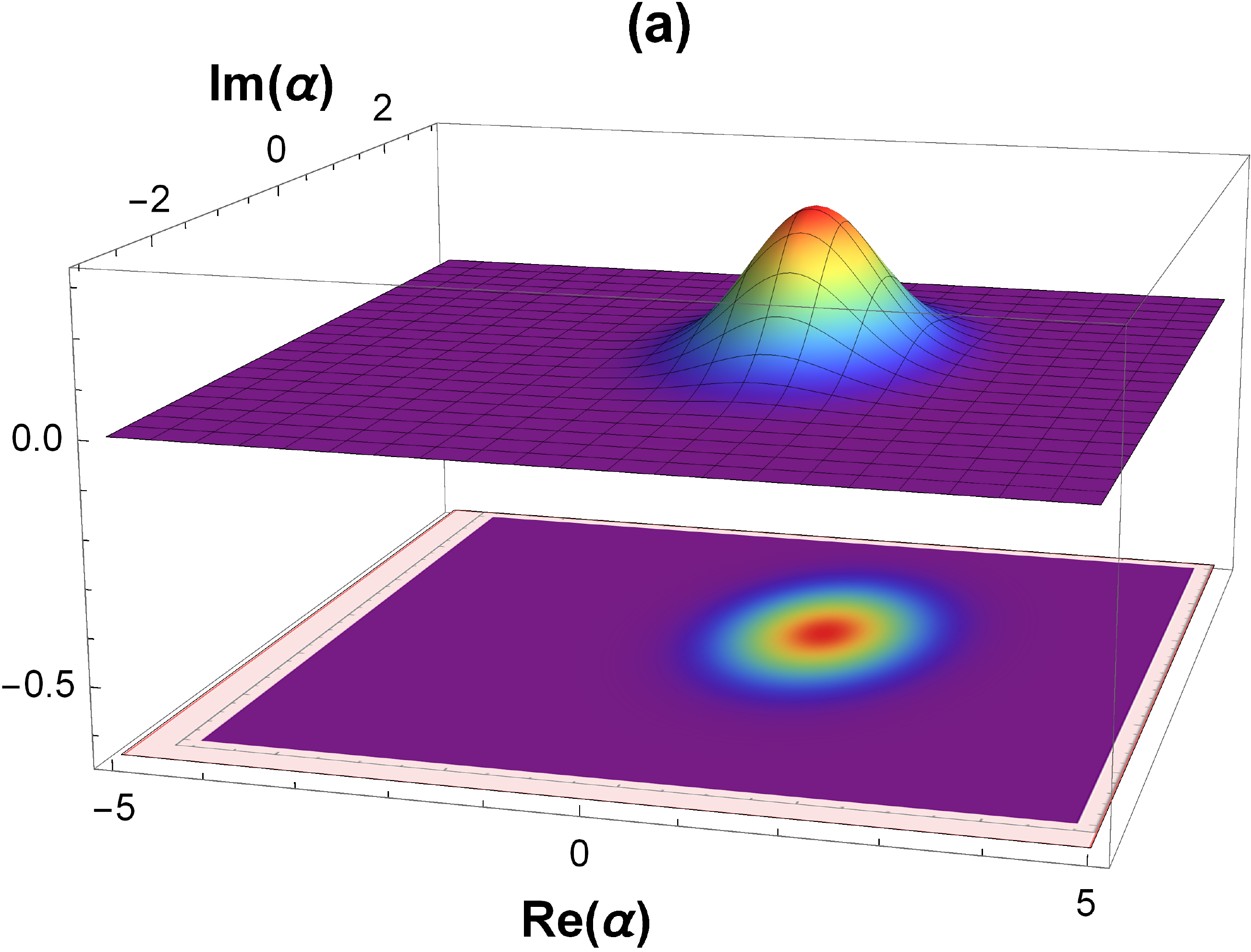}\includegraphics[width=6cm]{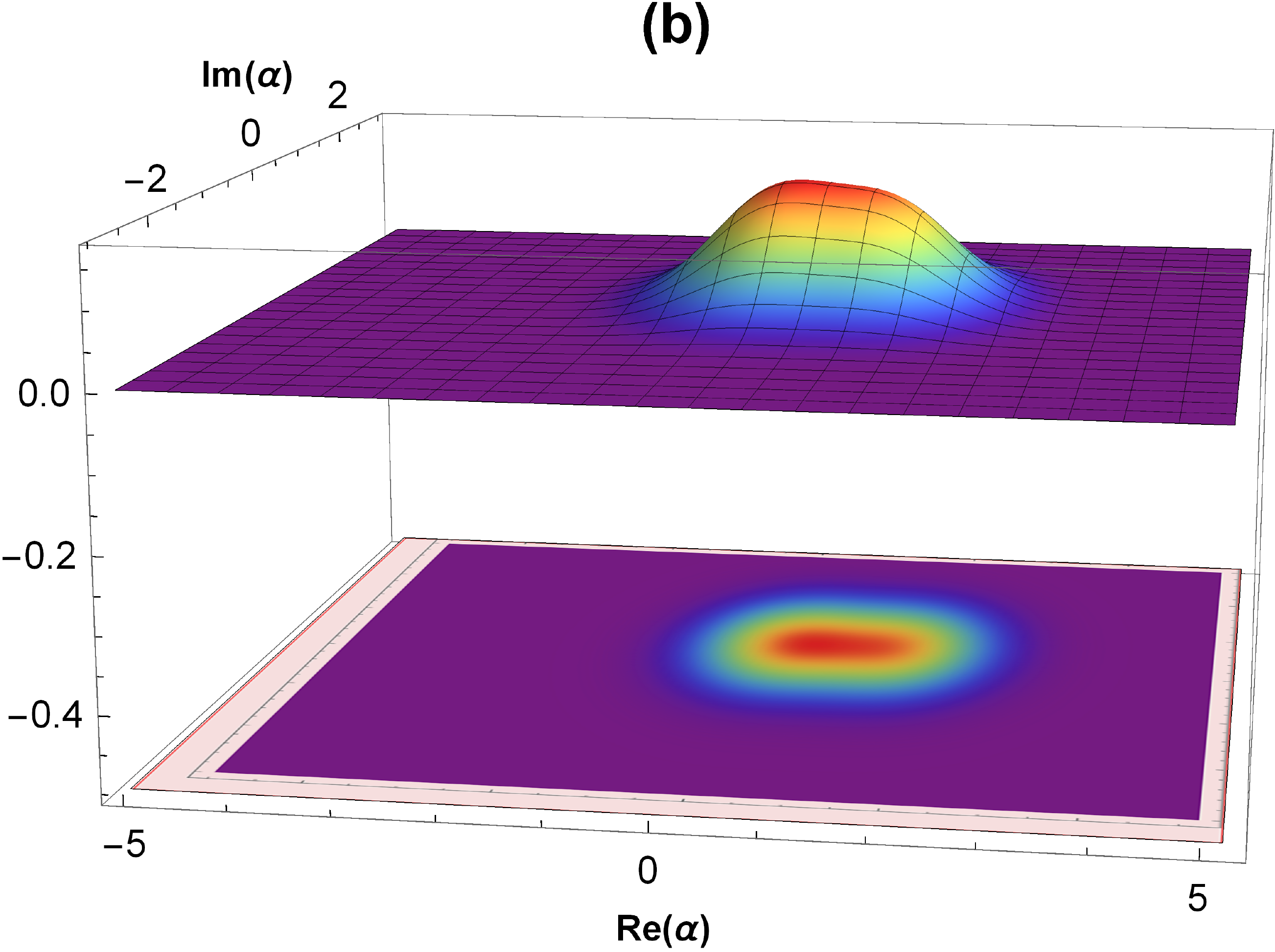}\includegraphics[width=6cm]{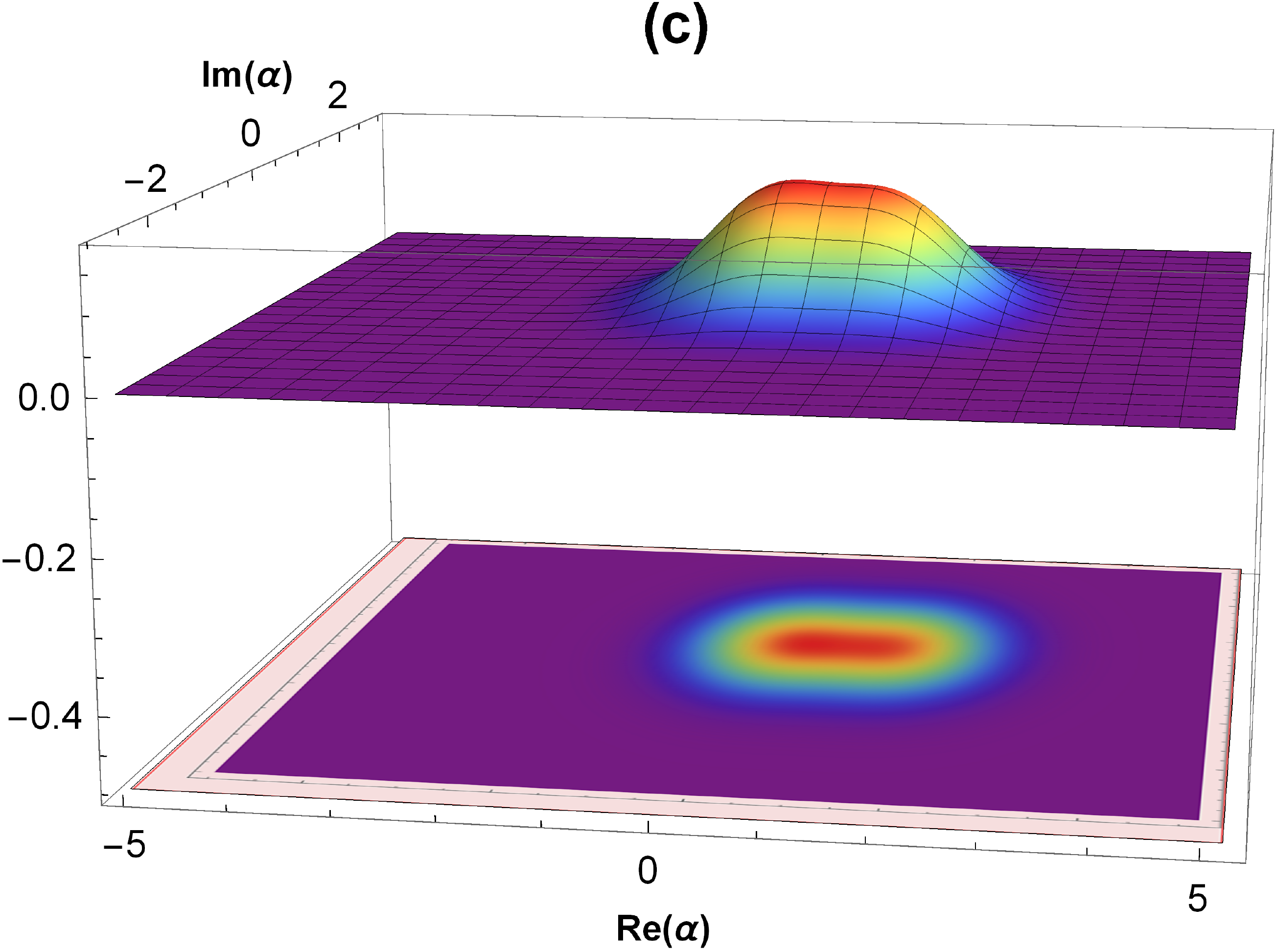}

\includegraphics[width=6cm]{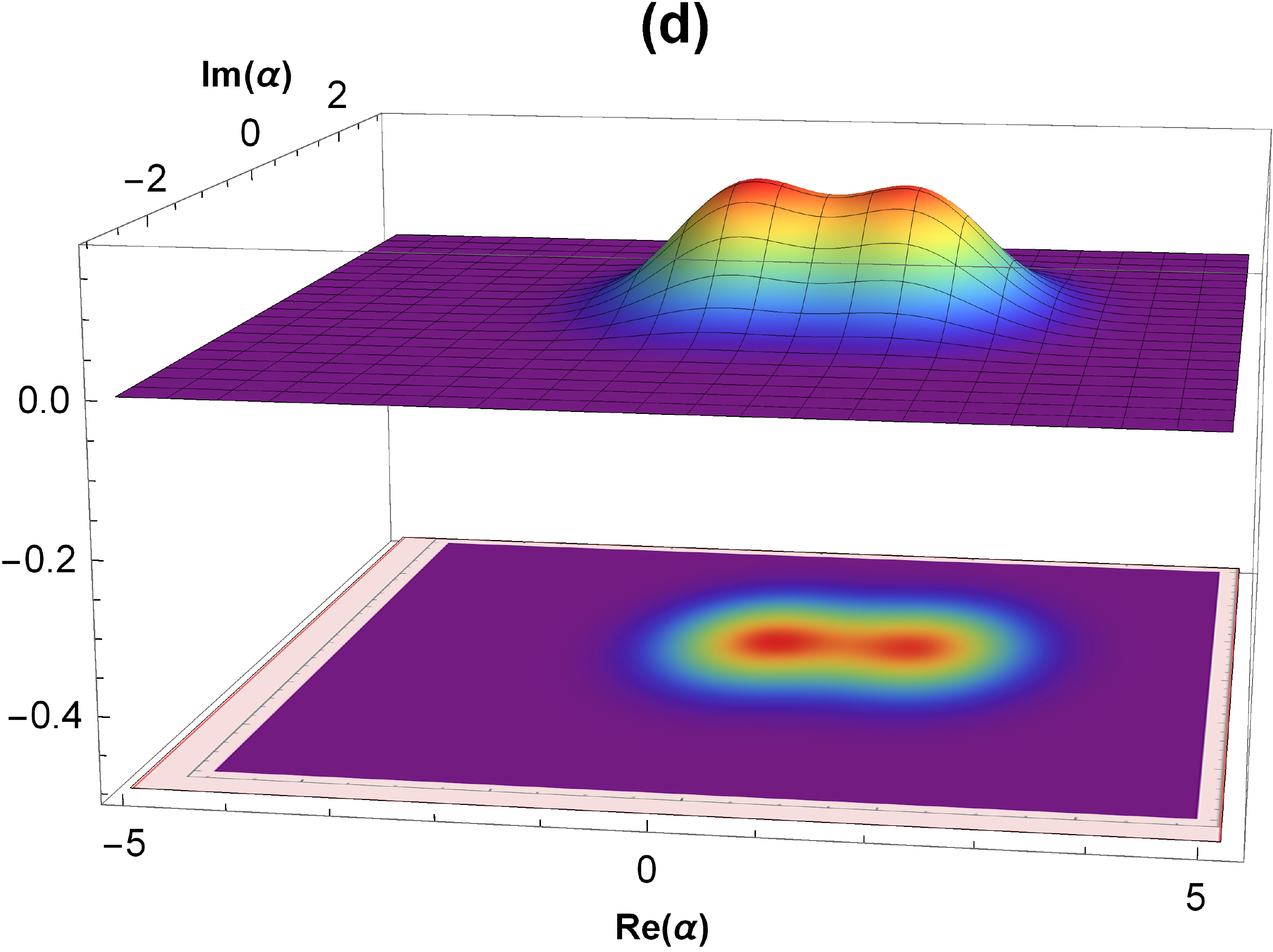}\includegraphics[width=6cm]{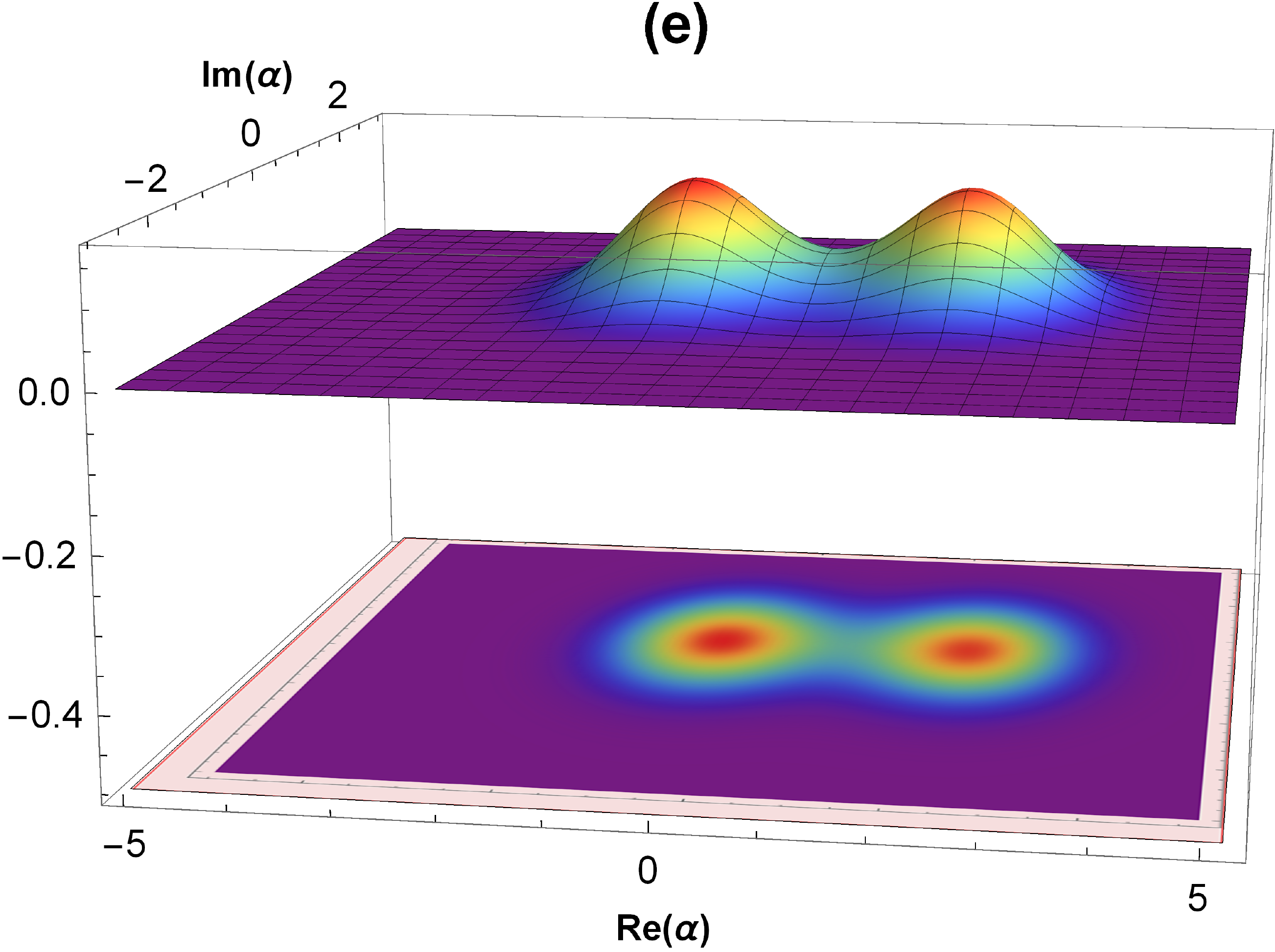}\includegraphics[width=6cm]{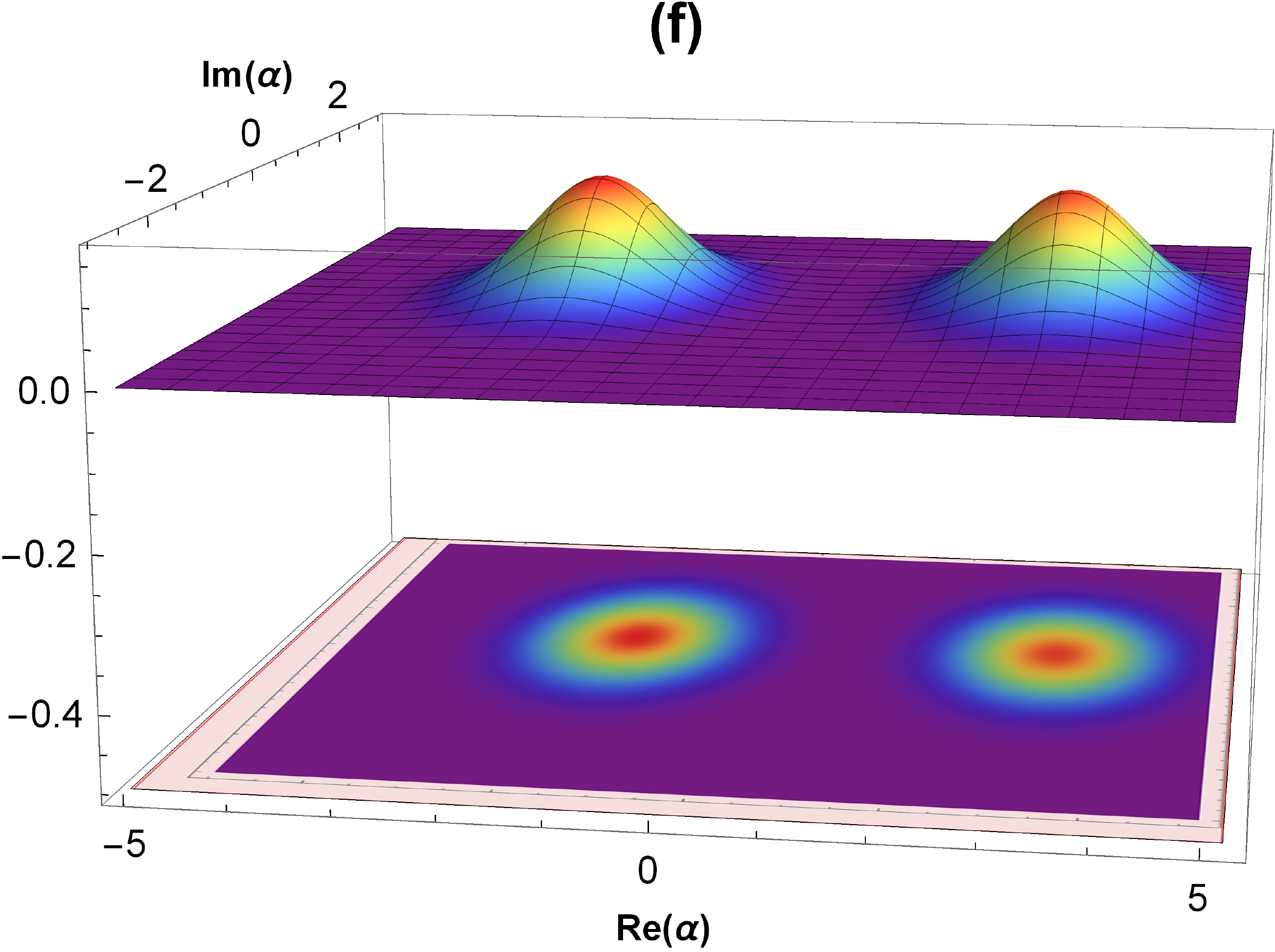}

\caption{\label{fig:3} $Q_{\Phi}(\alpha)$ function (see the Eq. (\ref{eq:60}))
for different coupling strengths $\Gamma$ : (a) no
interaction case ($\Gamma=0$); (b) $\Gamma=0.5$; (c) $\Gamma=1$;
(d) $\Gamma=2$; (e) $\Gamma=3$; (f) $\Gamma=5$. Here, we take $r=1$,
$\varphi=\frac{\pi}{6}$, and $\theta=0.01$.}
\end{figure}

\end{widetext}

In this subsection, in order to better visualize the weak-to-strong transition
for coherent pointer states, we calculate the phase space distribution---
$Q$ function. In quantum mechanics, Heisenberg's uncertainty
relations prevent the notion of a system being characterized by a
point in phase space. However, since coherent states minimize the
uncertainty product for two orthogonal quadrature operators and
the two uncertainties are equal, we can write
the phase space distribution of a quantum states by using the coherent
state representation. Notice that the quadrature operators are
dimensionless and scaled versions of the position and momentum operators.
There are three main phase space quasi-probability distributions
\citep{GerryIntroductory}, i.e. the Glauber-Sudarshan P-function,
the Husimi-Kano Q function, and the Wigner function. Among them, the Q
function is the expectation value of the density operator
on a coherent state. As shown in \citep{M1989} the interference
effect in phase space can be described by considering the properties
of Q function
\begin{equation}
Q_{\psi}(\alpha)=\frac{1}{\pi}\vert\langle\alpha\vert\psi\rangle\vert^{2}
\end{equation}
for the state $\vert\psi\rangle$ under study. This expression may
also be rewritten as
\begin{equation}
\langle\alpha\vert\psi\rangle=\sqrt{\pi Q_{\psi}(\alpha)}e^{i\phi_{\psi}(\alpha)}\,.
\end{equation}
If $\vert\alpha\rangle$ and $\vert\psi\rangle$ denote two
different quantum states, then their inner product $\langle\alpha\vert\psi\rangle$
is linked to the overlap in the phase space. In particular,
$\pi Q_{\psi}(\alpha)$ denotes the
overlap and $\phi_{\psi}(\alpha)$  is its phase. If we take
$\vert\psi\rangle=\vert\Psi^{\prime\prime}\rangle$, its $Q$ function
can be calculated as

\begin{align}
Q_{\Psi^{\prime\prime}}(\alpha) & =\frac{1}{\pi}\vert\langle\alpha\vert\Psi^{\prime\prime}\rangle\vert^{2}\nonumber \\
 & =\frac{1}{\pi}\frac{1}{1-\cos\left(2\theta\right)\cos\left(2\Gamma r\sin\varphi\right)e^{-\frac{1}{2}\Gamma^{2}}}\nonumber \\
 & \!\!\!\!\times\!\!\{\sin^{2}\left(\frac{\pi}{4}-\!\theta\right)e^{-\vert\alpha-\beta-\frac{\Gamma}{2}\vert^{2}}\text{\!\!\!}+\cos^{2}\left(\!\frac{\pi}{4}-\!\theta\right)e^{-\vert\alpha-\beta+\frac{\Gamma}{2}\vert^{2}}\nonumber \\
 & -\cos2\theta\cos\left(2\Gamma r\sin\varphi\right)e^{-\frac{1}{2}\vert\alpha-\beta-\frac{\Gamma}{2}\vert^{2}}e^{-\frac{1}{2}\vert\alpha-\beta+\frac{\Gamma}{2}\vert^{2}}\}.\label{eq:60}
\end{align}
The $Q$- function of our initial coherent pointer state $\vert\beta\rangle$
is given by
\begin{equation}
Q_{coh}(\alpha)=\frac{1}{\pi}\vert\langle\alpha\vert\beta\rangle\vert^{2}=\frac{1}{\pi}e^{-\vert\alpha-\beta\vert^{2}}.\label{eq:61}
\end{equation}
By comparing the Eqs. (\ref{eq:60}) and (\ref{eq:61}), we can see
that $Q_{\Psi^{\prime\prime}}(\alpha)$ contains three terms:
the first and second terms represent the $Q-$functions of two different
coherent states with amplitudes $\vert\beta+\frac{\Gamma}{2}\vert$
and $\vert\beta-\frac{\Gamma}{2}\vert$, respectively, whereas the third
one describes their interference. The extra
$\pm\frac{\Gamma}{2}$ in the exponential function part of the first
and second terms of Eq.(\ref{eq:60}) [compare to Eq. (\ref{eq:61})]
are caused by the interaction between the measured system and the
pointer. After the measurement, the wave-packet separates into
sub-wave-packets corresponding to the two eigenvalues
of the Pauli-$x$ operator. We also
notice that in this expression the transition factor
$e^{-\frac{1}{2}\Gamma^{2}}$ is still present. If we fix the
coherent state parameter $\beta$, with the increasing of the
 coupling parameter $\Gamma$, the overlap gradually decreases and
 the third term vanishes.
To describe the measurement transition phenomena in the phase-space,
we show, in Fig. \ref{fig:3}, the function
$Q_{\Phi}(\alpha)$ function
for different coupling strengths $\Gamma$. In panel
(a) we see the $Q$- function of the coherent state $\vert\beta\rangle$, which is
a Gaussian located at $\alpha_{r}=\sqrt{3}/2$ and $\alpha_{i}=1/2$.
This Gaussian function is symmetric, the  corresponding contour plot denotes
the $1/e$ decay of the $Q_{coh}(\alpha)$-function. If the interaction
is weak, there are two overlapping Gaussian wave-packets
and thus interference effects appear, due to the third term in Eq.(\ref{eq:60})
{[}see the Fig. \ref{fig:3} (b)-(e){]} . However, as it can be seen in
Fig. \ref{fig:3}(f), if the coupling between the system and pointer is strong
enough, the initially overlapping two sub-wave-packets are fully separated, and the measurement transition from weak to strong measurement
regimes occured. Here we take $\beta=\beta_{r}+i\beta_{i}=\sqrt{3}/2+0.5i$,
though the $Q$-function of a coherent state
is always a Gaussian fuction, independently on the value of the
parameters $\beta_{r}$ and $\beta_{i}$.
By comparing Fig.\ref{fig:3}(a) to Fig.\ref{fig:3}(e) and
(f), we find that the radius of the circular contour plot is indeed
independent of $\beta$. Since a coherent state corresponds to a
displaced ground state of a harmonic oscillator, the fluctuations
of the electric field operator in a coherent state are independent
of the displacement $\beta$, i.e. fluctuations are only determined
by the properties of the
harmonic oscillator, and not by the displacement amplitude.

\section{\label{sec:4}A possible experimental realization}
In this Section, we put forward a possible implementation of our
measurement scheme in a trapped-ion system. As shown in previous studies ,
{[}8{]}, trapped ions subject to laser cooling represent an excellent platform for
quantum state preparation and manipulation \citep{PhysRevLett.70.556}
since they are characterized by a decoherence time much longer than
the typical times required to perform quantum measurements. Furthermore,
the (quantum) motion state of trapped ions can be fully characterized by quantum tomographic measurements \citep{PhysRevLett.75.2932,PhysRevLett.77.4281}. Up
to date, a variety of schemes have been proposed and experimentally
observed for the preparation of nonclassical motional  states
of a trapped ion \citep{PhysRevLett.76.1796,PhysRevA.48.2393,033,2019Quantum},
such as Fock \citep{PhysRevLett.70.762}, coherent \citep{PhysRevLett.107.243902,PhysRevA.62.052108,RAFFA2012330,2016Generation},
squeezed \citep{PhysRevA.52.809,PhysRevLett.70.556,PhysRevA.101.052331},
Schrodinger cat \citep{PhysRevLett.76.608,PhysRevA.54.1532,PhysRevA.55.2478,Haroche2006,RevModPhys.85.1103},
pair coherent \citep{PhysRevA.54.R1014}, and pair cat \citep{PhysRevA.54.4315}
states. In order to implement our scheme, we consider an
ion trapped in a harmonic potential with frequency $\nu$, and driven
by two laser beams interacting resonantly with the system, tuned to
the lower (red) and upper (blue) vibrational sidebands, respectively.
By taking the Lamb-Dicke regime \citep{111} into account,
the total system Hamiltonian in the interaction picture  is given
by
\citep{PhysRevA.58.761,PhysRevLett.75.2932}
\begin{align}
H & =\eta\Omega\left(\hat{\sigma}_{x}\sin\phi_{+}+\hat{\sigma}_{y}\cos\phi_{+}\right)\nonumber \\
 & \otimes\left(\sigma\sin\phi_{-}\hat{P}-\frac{\hbar}{2\sigma}\cos\phi_{-}\hat{X}\right),\label{eq:62}
\end{align}
where $\eta$ is the Lamb-Dcike parameter, $\text{\ensuremath{\Omega}}$
is the Rabi frequency, and $\phi_{\pm}=\frac{1}{2}\left(\phi_{red}\pm\phi_{blue}\right)$
are phases depending on the lower and upper sideband lasers phases
$\phi_{red}$ and $\phi_{blue}$. Here, $\hat{X}=\sigma\left(\hat{a}+\hat{a}^{\dagger}\right)$
and $\hat{P}=\frac{i}{2\sigma}\left(\hat{a}^{\dagger}-\hat{a}\right)$
are the position and momentum operators of the vibrational degree of freedom of the ion,  and $\sigma=\sqrt{1/2\nu m}$ characterizes the size
of the motional state, which itself depends on the mass $m$ and on the
vibrational $\nu$ of the ion. Another important point is that here
the ion is considered as a two level system. In the above Hamiltonian,
the $\hat{\sigma}_{x}$ and $\hat{\sigma}_{y}$ are the Pauli-$x$
and -$y$ operatorsm which can be written in terms of ion's ground
($\vert\downarrow\rangle$) and optically excited ($\vert\uparrow\rangle$)
states as $\hat{\sigma}_{x}=\vert\uparrow\rangle\langle\downarrow\vert+\vert\downarrow\rangle\langle\uparrow\vert$
and $\hat{\sigma}_{y}=i\left(\vert\uparrow\rangle\langle\downarrow\vert-\vert\downarrow\rangle\langle\uparrow\vert\right)$,
respectively. The Lamb-Dcike parameter $\eta$ is related to the wave
vector $k$, and is given by$\eta=k/\sqrt{2\nu m}$. In the derivation
of the above Hamiltonian we have assumed $\eta\ll1$.

If we take the external and internal parts of the two level trapped
ion as the pointer and measured system, respectively, Eq. (\ref{eq:62})
describes the typical von Neumann type measurement by adjusting some
of the parameters. In general, the system observable $\hat{A}$ satisfies
$\hat{A}^{2}=\hat{A}$ or $\hat{A}^{2}=\hat{I}$, and the $\hat{A}^{2}=\hat{I}$
case is very usual in measurement problems. If we set $\phi_{-}=\frac{\pi}{2}$
, $\phi_{+}=\frac{\pi}{2}$ or $0$, the above interaction Hamiltonian
takes the form
\begin{equation}
H=g\hat{\sigma}_{x}\otimes\hat{P},\label{eq:63}
\end{equation}
or
\begin{equation}
H=g\hat{\sigma}_{y}\otimes\hat{P},\label{eq:64}
\end{equation}
respectively, with $g=\eta\sigma\Omega$. These correspond to the von
Nuemann type measurement Hamiltonians introduced in Sec. \ref{sec:2}.
In a recent study \citep{Nature2020}, the authors have investigated
 the weak-to-strong
measurement transition problem by taking a zero mean Gaussian state
as pointer in $^{40}Ca^{+}$ trapped ion system. Since state preparation
in trapped ion system is very realiable, and the generated states are stable and mantain coherence over a long time, we anticipate that our
general approach to weak-to-strong measurement transition may be
implemented by using the Hamiltonians in Eq. (\ref{eq:63}) or
Eq.(\ref{eq:64}) using Fock \citep{Jos2000,2016,2019,PhysRevResearch.2.023372},
SPACS \citep{033} and Schodinger cat pointer states \citep{PhysRevLett.116.140402}.

\section{\label{sec:5}Conclusion and remarks}

In conclusion, we have given a general expressions of the position and
momentum shifts which holds for any infinite-dimensional pointer states
and any value of coupling strengths between the measured system and the
pointer. At first, we have derived the general formula of the position shift
corresponding to postselected weak measurements {[}see
Eq. (\ref{eq:25}){]}, and have verified that it agrees with previous results.
We have also proved that by adjusting the coupling strength
parameter $\Gamma$, we can link the weak value and the conditional
expectation value of system observable obtained by the shifts of the
pointer in weak and strong measurement regimes, respectively. As a
typical example of our general approach, we have illustrated the weak-to-strong
measurement transition for a coherent pointer state, also analyzing the phenomenon in the phase space by using the $Q$- function.
Finally, we have discussed a possible experimental realization of
our weak-to-strong measurement transition scheme
in trapped ion system.
\begin{acknowledgments}
This work was supported by the National Natural Science Foundation of China (Grants No. 11865017), the Natural Science Foundation of Xinjiang Uyghur Autonomous Region of China (Grant No. 2020D01A72).

\end{acknowledgments}

\bibliographystyle{apsrev4-1}
\bibliography{ref}

\end{document}